\documentclass[osajnl,twocolumn,showpacs,superscriptaddress,10pt]{revtex4-1} 
\usepackage{amsmath,amssymb,graphicx}
\begin{document}

\title{Null-field radiationless sources: Comment
 }

\author{Iver Brevik}
\affiliation{Department of Energy and Process Engineering, Norwegian University of Science and Technology, 7491 Trondheim, Norway
}


\begin{abstract}
 We elaborate upon a recent paper of E. Hurwitz and G. Gbur [Opt. Lett. {\bf 39}, 6529 (2014)], dealing with null-field sources. After commenting on their theory in Minkowski space,  we show how the theory can be conveniently generalized to  the case of curvilinear space.
\end{abstract}


\maketitle 


The  recent interesting paper of Hurwitz and Gbur \cite{hurwitz14} deals with  nonradiating sources - sources whose fields are limited to a finite region of space and exactly equal to zero outside this region. The topic is in itself old, with roots going back to Ehrenfest (1910). Null-field sources have later been considered elsewhere, for instance in Ref.~\cite{nikolova05}. The interest in this topic is understandable, among other things because of its relationship to cloaking devices.

 We will focus on two issues related to Ref.~\cite{hurwitz14}:

 {\bf 1.} Consider  first Maxwell's equations for monochromatic modes in Minkowski space  ($\eta_{00}=-1, c=1$):   $ {\rm curl}\, {\bf E}=i\omega {\bf B}, \quad {\rm div}\, {\bf B}=0, \,
{\rm curl} \,{\bf H}=-i\omega {\bf D}, \quad {\rm div }\,{\bf D}=0.$
Using  $\bf D=E+P, B=H+M$ we obtain from this ${\rm curl}\, {\rm curl}\, {\bf E}-\omega^2 {\bf E}=\omega^2{\bf P}+i\omega \,{\rm curl}\, \bf M.$ Thus, if
\begin{equation}
-i\omega {\bf P}+{\rm curl}\, {\bf M}=0, \label{1}
\end{equation}
 the above equation for $\bf E$ becomes homogeneous. That this really yields a null-field solution, can be verified via Green-function methods \cite{hurwitz14}.

 The expression (\ref{1}), when written in a slightly more general form as $(\partial_0 {\bf P}+{\rm curl}\, \bf M)$, gives the total current density in matter when  extraneous currents are omitted. One may ask: can one herefrom conclude that the electromagnetic force density can written  simply as $ (\partial_0 {\bf P}+{\rm curl}\, \bf M)\times B$?  This is a very central point in electrodynamics. The argument seems quite natural formally, and  has occasionally been presented in the literature, the first one probably being Poincelot \cite{poincelot}. In our opinion the answer to the question is however no. Space forbids us to go into detail here, but the extensive investigation given in Ref.~\cite{brevik79} shows that all known experiments in optics are more naturally explained in terms of the Minkowski, or equivalently the Abraham, energy-momentum tensors. Cf. also Ref.~\cite{brevik14}. We emphasize that this conclusion rests primarily on {\it experimental}, not on theoretical, input.

{\bf 2.} Second, we  point out that the above theory can conveniently be generalized to the case of curvilinear space when the metric is time-independent and time-orthogonal. Then the four-dimensional metric $g_{\mu\nu}$ reduces to $g_{00}$ and $g_{ik}$, the nondiagonal components being zero.
We introduce the antisymmetric pseudotensor  $\epsilon_{ijk}=\gamma^{1/2}\delta_{ijk},  $
with $\gamma={\rm det}(g_{ik}),~ \delta_{123}=1$.  There are  two field tensors, $F_{\mu\nu}$ and $H^{\mu\nu}$,   related to the  electromagnetic fields via $F_{0i}=-E_i, \quad F_{ik}=\epsilon_{ikl}B^l,
H^{0i}=(-g_{00})^{-1/2}D^i, \quad H_i=\frac{1}{2}\epsilon_{ikl}H^{kl}. $ As the curvilinear space  functions like a "medium" with permittivity equal to the permeability: $\varepsilon=\mu=(-g_{00})^{-1/2}$, the constitutive relations take the form ${\bf D}=(-g_{00})^{-1/2}({\bf E+\bf P}), \quad {\bf B}=(-g_{00})^{-1/2}({\bf H +M})$, whereby the field equation for $\bf E$ becomes ${\rm curl}\,[(-g_{00})^{1/2}{\rm curl}\,{\bf E}]-\omega^2(-g_{00})^{-1/2}{\bf E}
=\omega^2 (-g_{00})^{-1/2}{\bf P} +i\omega\, {\rm curl}\,\bf M. $
The null-field condition analogous to Eq.~(\ref{1}) is
\begin{equation}
-i\omega (-g_{00})^{-1/2}{\bf P} + {\rm curl}\,{\bf M} =0. \label{2}
\end{equation}
A few examples can here be considered, the most typical one being  the Rindler space corresponding to a constant
 acceleration $a$ along the $x$ axis with respect to the inertial background space. (This space even allows a huge Casimir effect at finite temperature \cite{zhao11}.) The metric is  $ds^2=-a^2x^2dt^2+d{\bf r}^2$. Putting $a=1$ we obtain in this case the null-field condition  $-(i\omega/x){\bf P}+{\rm curl}\, {\bf M}=0.$ Other typical examples are the anisotropic Kasner space, and the Schwarzschild space.

 Note: this result concerns fundamental physics primarily. It shows the great validity of the electromagnetic formalism when generalized to curvilinear space, and is in general use in general relativity and cosmology. Under daily-life conditions the transformation technique has found an interesting, and perhaps unexpected, application in cloaking devices, as mentioned.





\end{document}